\documentclass[12pt,preprint]{aastex}

\slugcomment{} \shorttitle{Optical monitoring of 3C 390.3 and its
periodicity} \shortauthors{Tao, Fan, \& Qian}
\begin{document}

\title {Optical Monitoring of 3C 390.3 from 1995 to 2004 and Possible Periodicities in the Historical Light Curve}
\author{Jun Tao\altaffilmark{1,2}, Junhui Fan\altaffilmark{3,4},  Bochen Qian\altaffilmark{1,2},
and Yi Liu\altaffilmark{3}} \email{taojun@shao.ac.cn}
\altaffiltext{1}{Shanghai Astronomical Observatory, CAS, 80 Nandan
Road, Shanghai, 200030, China} \altaffiltext{2}{Joint Institute for
Galaxies and Cosmology, ShAO and USTC, CAS} \altaffiltext{3}{Center
for Astrophysics, Guangzhou University, Guangzhou 510006,
China}\altaffiltext{4}{Physics Institute, Hunan Normal University,
Changsha, China}

%Note change of English translation of institution name

\begin{abstract}

We report V, R, and I band CCD photometry of the radio galaxy 3C
390.3 obtained with the 1.56-m telescope of the Shanghai
Astronomical Observatory from March 1995 to August 2004. Combining
these data with data from the literature, we have constructed a
historical light curve from 1894 to 2004 and searched for
periodicities using the {\it CLEANest} program. We find possible
periods of $8.30\pm1.17$, $5.37\pm0.49$, $3.51\pm0.21$, and
$2.13\pm0.08$ years.
\end{abstract}


\section{Introduction}

Active galactic nuclei (AGNs) can be divided into two classes:
radio-loud and radio-quiet. A small subgroup of radio-loud AGNs have
displayed flux variability on time scales ranging from hours to
decades (see a review by Fan 2005). Examples of long-term variations
can be found in the light curves of 3C 120 (Jurkevich et al. 1971),
OJ 287 (Sillanp$\rm\ddot{a}\ddot{a}$ et al. 1988), 3C 345 (Kidger et
al. 1992), PKS 0735+178 (Pollock, 1975, Fan et al. 1997, Qian \& Tao
2004), BL Lacertae (Fan et al. 1998), 3C 66A (Lainela et al. 1999),
3C 279 (Fan 1999), S5 0716+714 (Qian et al. 2002), Mrk 335 (Tao et
al. 2004), PKS 1510-089 \& MA 0829+047 (Liller \& Liller 1975), B2
1101+38 (Miller 1975) and in various other radio-selected BL
Lacertae objects (Fan et al. 2002).

3C 390.3 (z=0.0561; Osterbrock et al. 1975) is a well-known
broad-line radio galaxy with broad double peaked emission line
profile (Sandage 1966; Lynds 1968). It is also an extended
double-lobed FR II radio source. Leahy \& Perley (1995) discussed
the jets and hotspots of it and they also provided some basic data
of this source. The VLBI observations at 5 GHz show evidence of
superluminal motions, with $v/c\approx4.0$ (Alef et al. 1996). 3C
390.3 has a well known history, both in the continuum light and in
emission lines(e.g., Dietrich et al. 1998; Shapovalova et al. 2001;
Sandage 1967, 1973; Zheng 1996). Prieto \& Kotilainen(1997) detected
optically-resolved emission coincident with the brightest region of
the northern radio lobe (hot spot B) in B, R, and I bands. Dunn et
al.(2006) provided a database of Ultraviolet continuum light curves
for AGNs, and 3C 390.3 is one of them. {\it ROSAT PSPC} of 3C 390.3
show resolved X-ray emission spatially located at the position of
the northern radio hotspot(Prieto 1997). It is also a highly
variable X-ray source (Inda et al. 1994; Eracleous et al. 1996;
Wozniak et al. 1998; Leighly et al. 1997; Leighly \& O'Brien 1997;
Gliozzi et al. 2003). Gaskell(2006) discussed the relationship
between the X-ray, UV, and optical variability in 3C 390.3. The
narrow-line region in 3C 390.3 is probably on the order of 1 lt yr,
and a significant part of the NLR is located near the line of
sight(Zheng et al. 1995). The black hole mass of 3C 390.3 has been
estimated through reverberation mapping (e.g., McLure \& Dunlop
2001; Botte et al. 2004; Shapovalova et al. 2001), ranging from
$2.2\times 10^{8}$ to $2.1 \times 2.1^{9} M_{\bigodot}$ .

3C 390.3 has been one of our primary targets for long-term
monitoring in the optical band at the Shanghai Astronomical
Observatory(SHAO). In this paper, we present V, R, and I data taken
in the period from 1995 to 2004. We also compile historical data
from 1894 to 2004. We obtain a total of 786 observational points, to
which we adopt the $CLEANest$ method to search for periodicities.

\section{Observations and data reduction}

The observations presented here were obtained with the 1.56-m
telescope at SHAO from 1995 March 6 to 2004 August 14. At first a
liquid nitrogen cooled CCD camera ($1024\times 1024$ pixels, 1 pixel
= 0.019 mm) was used. The field of view of this was 4.3 arcminutes
(1 pixel = 0.25 arcsec) when used directly at the Cassegrain focus,
and approximately 13 arcmins (1 pixel = 0.76 arcsec) when used with
a focal reducer. A new liquid nitrogen cooled CCD camera
($2048\times2048$ pixels, SITe CCD chip) has been used since August,
2002. The chip of this camera subtends 11 arcmin by 11 arcmin in
sky, with a scale is 0.31 arcsec per pixel (1 pixel = 0.024 mm).
Standard Johnson-Cousins V, R, and I filters were used. Typical
integration times were 300 s for I and R, and 600 s for filter V,
depending on sky conditions and the brightness of 3C 390.3. The flat
field images were taken at dusk and dawn. The bias images were taken
at the beginning and the end of the observations, while the
dark-field images were taken at the end. All observing data were
processed using the IRAF software package.

The seeing at the Sheshan Station of SHAO varied from 1.3 to 2.0
arcseconds FWHM and we have discussed this elsewhere (see Fig. 3 of
Qian et al. 2002). Since 3C 390.3 is not a point-like source, seeing
fluctuations may have an effect. Thus, we have followed the
recommendations of Cellone et al. (2000) and selected a photometric
aperture of radius 6.5 arcseconds to minimize potential
seeing-dependent effects. Cellone et al.(2000) gave the minimum
aperture radii in Table 2 of their paper, for a V=16 mag. AGN and
various host galaxy for magnitudes $V_{G}$. The fainter the $V_{G}$,
the smaller the photometric radii that can be used. Shapovalova et
al.(2001) used a 10 arcsec diameter photometric aperture,
corresponding a 5 arcsec radius aperture, which is also comparable
to the one we have used.  We used comparison stars A, B, and D of
Penston et al.(1971) and HST GS 4951:731.  V, R, and I magnitudes
were taken from Dietrich et al.(1998). Using differential photometry
with respect to each comparison star, $c_{i}, i = 1 - n$, within the
frame we obtain a magnitude $m_i$ for the object of interest. Then
the object's magnitude at that time is
\begin{equation}\label{}
\overline{m}=\frac{\sum m_{i}}{n}, \\
\end{equation}
where $n$ is the number of standard stars, in this case, 4. The
uncertainty $\sigma_{1}$ was calculated as
\begin{equation}\label{}
\sigma_{1}=\sqrt{\frac{\sum(m_{i}-\overline{m})^{2}}{n-1}}.\\
\end{equation}
The difference between comparison stars is
\begin{equation}\label{}\sigma_{2}=(m_{A}-m_{B})-\Delta m_{AB},
\end{equation}
where $\Delta m_{AB}$ is the magnitude difference of comparison
stars A and B as given by Dietrich et al.(1998), and $m_{A}$ and
$m_{B}$ are the observed magnitudes of comparison stars A and B. The
absolute value of $\sigma_2$ was used as an additional indicator of
the observational uncertainty.

The observational data are listed in Tables 1--3, in which the first
column gives the Julian date, the second column the magnitudes, the
third the $\sigma_{1}$ uncertainties, and the fourth the
$\sigma_{2}$ uncertainties.

\section{The light curves}

Figures 1 -- 3 show the light curves of 3C 390.3 and the magnitude
differences between comparison stars A and B in the V, R, and I
filters, respectively. In our observation period from March 1995 to
August 2004, variations of 1$^{m}$.372(15.023 to 16.395) in the V
band, 1$^{m}$.240(14.352 to 15.592) in the R band, and 1$^{m}$.284
(13.896 to 15.180) in the I band are found. The observations show
that 3C 390.3 was brightening from 1995 and reached its brightest in
1996 September. Then it slowly declined in brightness, but was
brightening again in September 2003.

We have compared our results with those of Dietrich et al.(1998).
Our data have 10 days of overlap with the data of Dietrich et al. We
converted their published flux data of 3C 390.3 to V magnitudes by
taking $V=-2.5\times log(f)+17.961$. For each day, the observational
data were averaged and they were shown in Fig. 4. This comparison
gave the relationship used to convert their fluxes to V magnitudes.

Dunn et al.(2006) determined the UV light curves of 3C 390.3 using
the Multimission Archive at STScI. The data were observed with the
International Ultraviolet Explorer, from 1978 to 1996. The light
curves of the three bands, centered at 1431, 1816, and 1912
Angstroms, all show a brightening trend from JD 2449750 to JD
2450162, which is consistent with our V, R, and I light curves for
the same period.

\section{The Historical Light Curve and Possible Periodicities}

\subsection{Historic Light Curve}
We have reconstructed the historical light curve by combining our
recent data with data from the literature (Cannon et al. 1971;
Selmes et al. 1975; Dietrich et al. 1998; Scott et al. 1976;
Sergeev et al. 2002; Lloyd 1984; Shapovalova et al. 2001; Sandage
1967, 1973; Shen et al. 1972; Yee \& Oke 1981; Pica et al. 1980,
1988). Most of these data set are for the V band or continuum near V
band. However, there are some B band data in the remaining papers.
In order to include these data, we needed to convert them to V band
magnitude. Shapovalova et al. (2001) provide simultaneous B and V
data (see Fig. 5). From their data we find that $V = 7.234 + 0.508
B$ with a standard deviation of 0.052 mag. This relationship was
then used to convert B-band magnitudes to V-band magnitudes to give
a historical light curve in the V band. This has a total of 786
observational points over a time interval of 111 years (from 1894 to
2004). We show the historical light curve in Fig. 6. It shows two
outbursts (1970 and 1996) and a possible third outburst (near 1939).

\subsection{Periodicity Analysis}

Using the historical light curve we have constructed for 3C 390.3,
we searched for possible periodicities.

As in our previous papers, we use the $CLEANest$ analysis (Foster
1995) to search for periodicities (Fan et al. 2006, 2007). The
$CLEANest$ analysis cleans spurious periodicities as follows. First,
the strongest single peak and corresponding aliases are subtracted
from the original spectrum, then the residual spectrum is scanned to
determine whether the next strongest remaining peak is statistically
significant. If so, then the original data are analyzed to find the
pair of frequencies which best models the data, these two peaks and
corresponding aliases are subtracted, and the residual spectrum is
scanned again. The process continues, producing successive
$CLEANest$ spectra, until all statistically significant frequencies
are included.

The light curve of 3C 390.3 shows that the data are unevenly sampled
with most data concentrated in the period from 1965 to 2004.
Observations before 1933 are particularly sparse. After using the
$CLEANest$ algorithm on the 1933-2004 data set we find that there
are seven independent frequency components required to ``clean'' the
 light curve. The $CLEANest$ spectrum is shown in Fig. 7. For
comparison, in Fig. 10 we show the results of using only the post
1965 data.

When analyzing unevenly sampled time series, the irregular spacing
introduces many complications into the Fourier transformations; it
can alter peak frequencies (slightly) and amplitudes (greatly), and
even introduce extremely large spurious peaks. Including all 3C
390.3 data points would not only give irregular sampling but would
also give undue weight to the most recent 40 yr of data. To obtain
regular sampling and to give more uniform weighting to different
epochs we binned the data from 1933 onwards. However, binning data
inevitably throws away information. To minimize this we adopted a
bin size of 0.02 yr (7.30 days) which is short enough compared to
the long-term periods we are looking for (years) and thus unlikely
to distort long-term variations. This binning gives us 361 points in
the binned light curve from 1933 to 2004.  In Fig. 8 we show the
$CLEANest$ spectrum for this binned light curve.

We also investigated the effect of removing a long-term linear trend
in the binned 1933--2004 historical light curve using $V(\rm
magnitude)=-0.849005+0.00820992 \times \rm t(year)$. In Fig.9 we
show the $CLEANest$ spectrum for this light curve.

Following Foster(1996), the variance of a frequency $Var(\omega)$
and the variance of the amplitude of the given frequency $Var(P)$
can be estimated by:
\begin{eqnarray}
    Var(\omega)&=&\frac{24\sigma_{res}^2}{NA^2T^2}\\
    Var(P)&=&\frac{2\sigma_{res}^2}{N}
\end{eqnarray}
where $\sigma_{res}$ is the variance of the residual data, $A$ is
the amplitude of the given frequency and $T$ is the total time span,
and $N$ means the number of data values in an observed time series.
The $\sigma_{res}^2$ is estimated by
\begin{equation}
    \sigma_{res}^2=\frac{NV_{res}}{N-3f-1},
\end{equation}
where $V_{res}$ is the variance of residual data,
$V_{res}=\left<res|res\right>-\left<1|res\right>$, and $f$ is the
number of discrete frequencies. $\left|x\right>$ represent a vector
in a N-dimensional vector space.
\begin{equation}
   \left|x\right>=[x(t_1),x(t_2),\dots,x(t_N)]
\end{equation}
   Defining the inner product of two vectors, $\left|f\right>$ and $\left|g\right>$, as
   the average value of the product $fg$ over the sampling ${t_{\alpha}}$
\begin{equation}
   \left<f|g\right> = \frac{\sum_{\alpha=1}^N f(t_{\alpha})g(t_{\alpha})}{N}
\end{equation}
 Defining the constant vector $\left|1\right>$ as
\begin{equation}
 \ \left|1\right>=[1,1,\dots,1] .
\end{equation}
 The variation of a vector $\left|f\right>$ is
 \begin{equation}
 \ V_f = \left<f|f\right> - \left<1|f\right>^2 \
\end{equation}

The results of these four $CLEANest$ analyses are listed in Table 4.
In this table we also give the False Alarm Probability (FAP) of each
of the $CLEANest$ frequency components, which depends on the
amplitude. Small FAP values support the reality of these
periodicities (see Fan et al. 2006 for details).

The unbinned V-band data, the binned data, the linear-trend removed
data, and the post-1965 data (see Table 4) all show that there are
apparent periodicities. Those matched periods derived from the first
three data sets are: $13298\pm6298$ days($15268\pm10857$ days),
$3700\pm 413$ days ($3296\pm 387$ days, $3026\pm 427$ days),
$1871\pm 106$ days ($1759\pm110$ days, $1961\pm 179$ days), $1297\pm
51$ days($1282\pm 77$ days), and $775\pm 18$ days ($775\pm 21$ days,
$776\pm 28$ days). However, since the period $13298\pm6298$
days($15268\pm10857$ days) is almost the length of the data
coverage, the period is not a physically significant one. From Fig.
6, we can see that most data was sampled in the period of 1965 to
2004(about 40 years). Therefore, we reinvestigated periods based on
the post 1965 data. The matched periods derived from the V band
data(1933-2004) and the post 1965 data(1965-2004) are: $7022\pm
1486$ days($7203\pm1564$ days), $3700\pm 413$ days($3660\pm404$
days), $1297\pm 51$ days($1306\pm51 $days), $847\pm22$
days($847\pm22$ days), and $238\pm2$ days($238\pm2$ days). In Table
4, we also present the phase of the corresponding period. Phase and
period can be used to identify a signal in multi-data set. We used
the obtained components to fit all the light curves and the results
are shown in Fig. 11.

According to the data processing theory, the results derived from
the linear-trend removed data set is more reliable. So the possible
periods are:
 $8.30\pm1.17$,
 $5.37\pm0.49$,
 $3.51\pm0.21$, and
 $2.13\pm0.08$ years.

\section{Discussion and Conclusions}

AGNs are variable throughout the whole electromagnetic spectrum. In
our monitoring program, we have observed the galaxy 3C 390.3 from
March 1995 to August 2004. The observations clearly show that the
source is variable in the optical band with the variation amplitude
of 1$^{m}$.372, 1$^{m}$.240, and 1$^{m}$.284 in the V, R, and I
bands respectively.  The V band historical light curve is compiled,
which has a time span of 111 years. Possible periods of
 $8.30\pm1.17$,
 $5.37\pm0.49$,
 $3.51\pm0.21$, and
 $2.13\pm0.08$ years
were found in the light curve by means of $CLEANest$ method.

It has been suggested that the long-term periodic outbursts of OJ
287 may be explained by a binary black hole
(Sillanp$\rm\ddot{a}\ddot{a}$ et al. 1988) or such outbursts may be
due to thermal and viscous instabilities in a thin accretion disk
(Meyer \& Meyer-Hofmeister 1984; Horiuchi \& Kato 1990). The
historical light curve of 3C 390.3 shows strong variability. Several
possible periods were found by $CLEANest$ method. The multiple
periods we derived may imply the instabilities in the disk. The
analysis of spectra of 3C 390.3 covering a period of over 20 yr may
indicate the binary black hole model(Gaskell 1996). Shapovalova et
al.(2001) have monitored this object between 1995 and 2000. Their
results do not support either the models of outflowing biconical gas
streams or those of supermassive binary black holes. They conclude
that they favor the accretion disk model.

\acknowledgements This work is partially supported by the Joint
Laboratory for Optical Astronomy of Chinese Academy of Sciences, the
National Natural Science Foundation of China (10573005,10633010),
the 973 project (No. 2007CB815405), and Science \& Technology
Commission of Shanghai Municipality (06DZ22101). We thank Martin
Gaskell for editing the English and for comments. We are grateful to
the referee, Dr. Paul Viita for his valuable comments.

\clearpage

\begin{deluxetable}{cccc}\tablewidth{0pt} \tablecaption{V-band magnitudes of 3C 390.3}
\tablehead{\colhead{Date(JD - 2449000)} & \colhead{Magnitude} &
\colhead{${\sigma}_{1}$} & \colhead{${\sigma}_{2}$}}\startdata
   783.3201 &  15.658 &  0.010 &  -0.004 \\
   783.3259 &  15.503 &  0.014 &  -0.016 \\
   783.3349 &  15.589 &  0.010 &   0.004 \\
   802.3073 &  15.696 &  0.023 &  -0.055 \\
   802.3107 &  15.450 &  0.014 &   0.003 \\
   803.3255 &  15.576 &  0.028 &  -0.055 \\
   845.1668 &  15.383 &  0.007 &   0.001 \\
   845.1734 &  15.594 &  0.004 &   0.000 \\
   924.0950 &  15.513 &  0.026 &   0.041 \\
   924.1245 &  15.598 &  0.019 &   0.028 \\
   935.0480 &  15.585 &  0.038 &  -0.051 \\
   935.0582 &  15.539 &  0.013 &  -0.001 \\
   935.0627 &  15.492 &  0.028 &  -0.040 \\
   943.0259 &  15.470 &  0.015 &  -0.025 \\
   943.0345 &  15.461 &  0.025 &  -0.047 \\
   947.0802 &  15.330 &  0.025 &   0.049 \\
   947.0870 &  15.436 &  0.025 &   0.049 \\
   977.9790 &  15.310 &  0.024 &   0.043 \\
   977.9976 &  15.320 &  0.043 &   0.087 \\
   978.0020 &  15.322 &  0.053 &   0.072 \\
  1196.2712 &  15.114 &  0.093 &   0.031 \\
  1196.2934 &  15.332 &  0.032 &   0.047 \\
  1197.2786 &  15.303 &  0.048 &  -0.053 \\
  1197.2850 &  15.480 &  0.041 &  -0.001 \\
  1199.2778 &  15.445 &  0.036 &   0.032 \\
  1199.2833 &  15.146 &  0.008 &   0.012 \\
  1221.1243 &  15.050 &  0.035 &  -0.047 \\
  1257.0869 &  15.328 &  0.044 &  -0.023 \\
  1257.0953 &  15.157 &  0.063 &   0.053 \\
  1328.0242 &  15.119 &  0.053 &  -0.075 \\
  1328.0316 &  15.382 &  0.047 &   0.047 \\
  1328.0402 &  15.047 &  0.073 &  -0.020 \\
  1334.0357 &  15.046 &  0.040 &  -0.014 \\
  1334.0427 &  15.056 &  0.035 &  -0.048 \\
  1371.9932 &  15.681 &  0.043 &   0.024 \\
  1372.0024 &  15.367 &  0.081 &   0.060 \\
  1550.3632 &  15.498 &  0.038 &   0.081 \\
  1577.2818 &  15.386 &  0.018 &  -0.040 \\
  1577.2894 &  15.408 &  0.041 &  -0.072 \\
  1577.2964 &  15.521 &  0.040 &  -0.081 \\
  1577.3030 &  15.579 &  0.041 &  -0.068 \\
  1586.2854 &  15.883 &  0.009 &   0.001 \\
  1586.3046 &  15.602 &  0.021 &  -0.005 \\
  1594.2563 &  15.361 &  0.059 &  -0.073 \\
  1594.2604 &  15.388 &  0.029 &   0.055 \\
  1594.2646 &  15.645 &  0.026 &  -0.052 \\
  1599.1615 &  15.570 &  0.017 &   0.019 \\
  1614.2056 &  15.515 &  0.035 &  -0.080 \\
  1614.2110 &  15.591 &  0.032 &   0.057 \\
  1614.2210 &  15.427 &  0.098 &  -0.034 \\
  1616.1638 &  15.764 &  0.027 &  -0.002 \\
  1616.1776 &  15.810 &  0.031 &  -0.015 \\
  1660.1593 &  15.590 &  0.074 &  -0.077 \\
  1660.1655 &  15.823 &  0.053 &   0.095 \\
  1692.0016 &  15.633 &  0.030 &  -0.017 \\
  1692.0042 &  15.741 &  0.024 &  -0.031 \\
  1692.0264 &  15.562 &  0.035 &  -0.061 \\
  1692.0333 &  15.432 &  0.024 &   0.019 \\
  1701.0160 &  15.731 &  0.024 &  -0.032 \\
  1701.0196 &  15.945 &  0.019 &  -0.011 \\
  2229.3530 &  15.518 &  0.027 &   0.004 \\
  2318.3192 &  15.853 &  0.045 &  -0.003 \\
  2348.1668 &  16.380 &  0.090 &   0.087 \\
  2348.1746 &  16.260 &  0.081 &   0.010 \\
  2437.0529 &  16.395 &  0.078 &   0.110 \\
  3041.2721 &  15.430 &  0.083 &  -0.118 \\
  3041.2866 &  15.673 &  0.087 &  -0.124 \\
  3094.2237 &  15.272 &  0.065 &   0.092 \\
  3094.2386 &  15.306 &  0.103 &   0.054 \\
  3100.1232 &  15.489 &  0.042 &   0.001 \\
  3112.3400 &  15.581 &  0.093 &   0.132 \\
  3518.0847 &  15.888 &  0.069 &  -0.023 \\
  3856.0297 &  15.242 &  0.063 &  -0.045 \\
  3856.0446 &  15.170 &  0.059 &  -0.083 \\
  3888.0554 &  15.023 &  0.068 &   0.096 \\
  3888.0695 &  15.733 &  0.069 &   0.097 \\
  4231.7777 &  16.060 &  0.060 &  -0.063 \\
\enddata
\end{deluxetable}
\begin{deluxetable}{cccc}\tablewidth{0pt} \tablecaption{R-band magnitudes of 3C 390.3}
\tablehead{\colhead{Date(JD - 2449000)} & \colhead{Magnitude} &
\colhead{${\sigma}_{1}$} & \colhead{${\sigma}_{2}$}}\startdata
   802.2993 &  15.206 &  0.052 &   0.073  \\
   802.3039 &  15.322 &  0.038 &  -0.025  \\
   803.3159 &  15.270 &  0.018 &  -0.003  \\
   819.0507 &  15.265 &  0.015 &   0.034  \\
   824.1183 &  15.234 &  0.040 &   0.049  \\
   845.1945 &  15.111 &  0.014 &   0.019  \\
   919.0410 &  15.305 &  0.028 &   0.052  \\
   924.0838 &  15.288 &  0.022 &   0.027  \\
   935.0480 &  15.149 &  0.015 &   0.027  \\
   943.0036 &  15.099 &  0.014 &  -0.005  \\
   943.0098 &  15.098 &  0.014 &  -0.004  \\
   947.0513 &  15.191 &  0.037 &   0.032  \\
   947.0587 &  15.154 &  0.035 &   0.066  \\
   977.9743 &  15.027 &  0.003 &  -0.004  \\
   977.9790 &  15.033 &  0.007 &  -0.010  \\
   977.9834 &  15.023 &  0.020 &   0.013  \\
  1196.2656 &  14.660 &  0.021 &   0.021  \\
  1196.2876 &  14.608 &  0.040 &   0.021  \\
  1197.2695 &  14.698 &  0.019 &   0.003  \\
  1197.2740 &  14.661 &  0.024 &  -0.029  \\
  1199.2660 &  14.820 &  0.035 &   0.041  \\
  1199.2722 &  14.775 &  0.044 &   0.056  \\
  1221.0993 &  14.769 &  0.024 &   0.035  \\
  1257.0738 &  14.880 &  0.054 &   0.034  \\
  1257.0799 &  14.871 &  0.045 &  -0.052  \\
  1328.0826 &  14.352 &  0.019 &   0.028  \\
  1328.0873 &  14.402 &  0.082 &   0.048  \\
  1328.0929 &  14.504 &  0.060 &   0.003  \\
  1332.0358 &  14.944 &  0.067 &  -0.095  \\
  1371.9859 &  14.656 &  0.062 &   0.030  \\
  1371.9889 &  14.718 &  0.086 &   0.031  \\
  1550.3524 &  15.144 &  0.012 &  -0.027  \\
  1550.3569 &  15.083 &  0.020 &   0.028  \\
  1577.3114 &  15.083 &  0.045 &   0.056  \\
  1577.3167 &  15.029 &  0.039 &   0.061  \\
  1577.3217 &  15.078 &  0.034 &   0.052  \\
  1577.3267 &  14.978 &  0.040 &   0.059  \\
  1577.3330 &  14.990 &  0.039 &   0.062  \\
  1586.2576 &  14.828 &  0.046 &   0.069  \\
  1586.2617 &  15.274 &  0.032 &   0.038  \\
  1586.2704 &  15.169 &  0.020 &   0.007  \\
  1586.2747 &  15.171 &  0.023 &   0.022  \\
  1586.2793 &  15.213 &  0.042 &   0.069  \\
  1594.2340 &  15.122 &  0.042 &   0.065  \\
  1594.2422 &  14.994 &  0.024 &  -0.012  \\
  1594.2447 &  15.196 &  0.024 &   0.019  \\
  1599.1528 &  15.192 &  0.047 &   0.060  \\
  1614.2360 &  15.219 &  0.015 &   0.019  \\
  1614.2418 &  15.191 &  0.016 &   0.016  \\
  1616.1541 &  15.138 &  0.020 &   0.029  \\
  1616.1572 &  15.211 &  0.015 &   0.021  \\
  1616.1600 &  15.271 &  0.030 &   0.056  \\
  1660.1492 &  15.339 &  0.034 &   0.008  \\
  1660.1524 &  15.210 &  0.021 &   0.012  \\
  1691.9937 &  15.205 &  0.014 &  -0.015  \\
  1691.9959 &  15.198 &  0.012 &   0.020  \\
  1691.9981 &  15.217 &  0.018 &  -0.005  \\
  1692.0182 &  15.181 &  0.014 &   0.027  \\
  1692.0209 &  15.170 &  0.027 &   0.032  \\
  1692.0234 &  15.223 &  0.030 &   0.005  \\
  1699.1292 &  15.067 &  0.029 &   0.009  \\
  1699.1339 &  15.181 &  0.034 &   0.045  \\
  1701.0020 &  15.317 &  0.029 &   0.024  \\
  1701.0053 &  15.301 &  0.007 &   0.013  \\
  1701.0087 &  15.373 &  0.006 &   0.003  \\
  1701.0122 &  15.380 &  0.029 &  -0.042  \\
  2229.3450 &  15.099 &  0.034 &   0.018  \\
  2318.3073 &  14.846 &  0.034 &   0.049  \\
  2348.1569 &  15.211 &  0.037 &   0.053  \\
  2348.1621 &  15.198 &  0.044 &   0.062  \\
  2437.0420 &  15.273 &  0.049 &   0.070  \\
  2757.3133 &  15.010 &  0.057 &   0.080  \\
  3017.2940 &  14.905 &  0.045 &  -0.051  \\
  3017.3345 &  15.014 &  0.049 &  -0.070  \\
  3018.3446 &  14.897 &  0.035 &  -0.049  \\
  3041.2560 &  14.911 &  0.071 &   0.062  \\
  3041.2636 &  14.955 &  0.038 &   0.078  \\
  3091.1142 &  14.839 &  0.087 &   0.022  \\
  3091.1253 &  15.106 &  0.066 &   0.076  \\
  3093.2495 &  14.950 &  0.057 &   0.026  \\
  3094.2063 &  14.977 &  0.036 &   0.027  \\
  3094.2139 &  14.991 &  0.052 &  -0.024  \\
  3100.1131 &  15.085 &  0.082 &  -0.023  \\
  3108.1328 &  15.135 &  0.071 &  -0.078  \\
  3108.1425 &  15.108 &  0.005 &   0.002  \\
  3112.3128 &  15.127 &  0.063 &  -0.040  \\
  3124.1414 &  15.023 &  0.021 &   0.012  \\
  3124.1552 &  15.033 &  0.079 &   0.010  \\
  3518.0390 &  15.211 &  0.119 &   0.168 \\
  3518.0766 &  15.329 &  0.099 &  -0.140 \\
  3519.0071 &  14.841 &  0.047 &   0.066 \\
  3519.0510 &  14.828 &  0.068 &   0.096 \\
  3856.0247 &  15.130 &  0.016 &   0.023 \\
  3856.0399 &  15.000 &  0.028 &   0.040 \\
  3873.9964 &  15.075 &  0.052 &   0.074 \\
  3887.9728 &  14.451 &  0.069 &  -0.098 \\
  3887.9856 &  15.090 &  0.144 &  -0.204 \\
  3887.9983 &  14.409 &  0.053 &   0.045 \\
  3888.0113 &  14.985 &  0.052 &   0.074 \\
  4231.7682 &  15.592 &  0.098 &  -0.139 \\
\enddata
\end{deluxetable}
\begin{deluxetable}{cccc}\tablewidth{0pt} \tablecaption{I-band magnitudes of 3C 390.3}
\tablehead{\colhead{Date(JD 2449000+)} & \colhead{Magnitude} &
\colhead{${\sigma}_{1}$} & \colhead{${\sigma}_{2}$}}\startdata
   783.3399 &  14.635 &  0.019 &  -0.036 \\
   783.3429 &  14.627 &  0.028 &  -0.008 \\
   783.3463 &  14.719 &  0.021 &  -0.038 \\
   783.3556 &  14.696 &  0.022 &  -0.044 \\
   802.3145 &  14.470 &  0.014 &   0.002 \\
   803.3013 &  14.600 &  0.019 &  -0.029 \\
   803.3078 &  14.551 &  0.019 &  -0.023 \\
   845.2020 &  14.532 &  0.023 &   0.004 \\
   845.2187 &  14.575 &  0.028 &   0.017 \\
   845.2234 &  14.566 &  0.030 &   0.030 \\
   919.0351 &  14.645 &  0.041 &   0.023 \\
   919.0451 &  14.629 &  0.038 &  -0.031 \\
   924.0796 &  14.630 &  0.045 &  -0.025 \\
   924.1049 &  14.690 &  0.056 &   0.020 \\
   934.9925 &  14.567 &  0.028 &  -0.043 \\
   935.0350 &  14.505 &  0.040 &  -0.076 \\
   935.0395 &  14.491 &  0.041 &  -0.036 \\
   942.9844 &  14.568 &  0.022 &  -0.018 \\
   942.9920 &  14.536 &  0.015 &  -0.017 \\
   947.0385 &  14.547 &  0.065 &  -0.001 \\
   947.0445 &  14.623 &  0.066 &  -0.001 \\
   961.9682 &  14.409 &  0.035 &  -0.017 \\
   961.9741 &  14.452 &  0.054 &  -0.032 \\
   961.9825 &  14.431 &  0.037 &  -0.012 \\
   977.9529 &  14.374 &  0.041 &   0.003 \\
   977.9578 &  14.432 &  0.039 &  -0.032 \\
   977.9626 &  14.389 &  0.092 &  -0.184 \\
  1196.2602 &  14.105 &  0.023 &  -0.013 \\
  1196.2785 &  14.096 &  0.039 &  -0.092 \\
  1197.2588 &  14.081 &  0.033 &   0.036 \\
  1197.2646 &  14.223 &  0.032 &   0.044 \\
  1199.2347 &  14.287 &  0.016 &  -0.004 \\
  1199.2490 &  14.219 &  0.025 &   0.000 \\
  1199.2556 &  14.220 &  0.022 &  -0.026 \\
  1221.0931 &  14.098 &  0.028 &   0.034 \\
  1257.0579 &  14.292 &  0.019 &  -0.007 \\
  1257.0645 &  14.238 &  0.031 &   0.004 \\
  1371.9801 &  14.117 &  0.028 &  -0.042 \\
  1550.3434 &  14.394 &  0.018 &   0.026 \\
  1550.3469 &  14.346 &  0.037 &   0.053 \\
  1577.2501 &  14.452 &  0.009 &   0.004 \\
  1577.2542 &  14.439 &  0.022 &   0.026 \\
  1577.2606 &  14.537 &  0.024 &   0.004 \\
  1577.2678 &  14.491 &  0.024 &  -0.012 \\
  1577.2715 &  14.469 &  0.025 &   0.013 \\
  1578.2352 &  14.378 &  0.040 &   0.035 \\
  1578.2389 &  14.553 &  0.065 &   0.096 \\
  1578.2842 &  14.512 &  0.047 &   0.019 \\
  1578.2877 &  14.503 &  0.069 &   0.075 \\
  1586.3170 &  14.578 &  0.014 &  -0.021 \\
  1586.3204 &  14.586 &  0.025 &   0.016 \\
  1586.3255 &  14.575 &  0.027 &   0.035 \\
  1594.2498 &  14.530 &  0.030 &   0.041 \\
  1594.2519 &  14.508 &  0.021 &   0.011 \\
  1594.2536 &  14.519 &  0.036 &   0.040 \\
  1599.1410 &  14.556 &  0.040 &   0.039 \\
  1599.1462 &  14.559 &  0.030 &   0.048 \\
  1614.1925 &  14.625 &  0.040 &  -0.085 \\
  1614.1991 &  14.602 &  0.031 &  -0.065 \\
  1614.2020 &  14.578 &  0.032 &  -0.062 \\
  1616.1458 &  14.620 &  0.013 &   0.011 \\
  1616.1483 &  14.536 &  0.092 &   0.015 \\
  1616.1509 &  14.528 &  0.091 &   0.033 \\
  1660.1375 &  14.905 &  0.041 &  -0.027 \\
  1660.1424 &  14.353 &  0.060 &   0.011 \\
  1660.1455 &  14.488 &  0.027 &  -0.013 \\
  1691.9837 &  14.575 &  0.028 &  -0.058 \\
  1691.9864 &  14.474 &  0.020 &  -0.007 \\
  1691.9895 &  14.543 &  0.033 &  -0.073 \\
  1692.0115 &  14.525 &  0.042 &   0.034 \\
  1692.0139 &  14.425 &  0.021 &   0.030 \\
  1692.0157 &  14.386 &  0.015 &  -0.006 \\
  1699.1103 &  14.432 &  0.009 &   0.009 \\
  1699.1143 &  14.562 &  0.013 &   0.017 \\
  1699.1182 &  14.510 &  0.013 &   0.009 \\
  1699.1234 &  14.268 &  0.035 &  -0.070 \\
  1700.9910 &  14.760 &  0.004 &   0.007 \\
  1700.9941 &  14.630 &  0.014 &  -0.027 \\
  1700.9972 &  14.784 &  0.035 &  -0.055 \\
  1700.9992 &  14.838 &  0.015 &  -0.010 \\
  2088.9659 &  14.368 &  0.029 &  -0.051 \\
  2229.3354 &  14.606 &  0.011 &   0.000 \\
  2229.3390 &  14.587 &  0.003 &   0.001 \\
  2318.2813 &  14.472 &  0.038 &   0.037 \\
  2318.2947 &  14.514 &  0.049 &   0.028 \\
  2348.1474 &  14.723 &  0.023 &   0.012 \\
  2348.1517 &  14.712 &  0.025 &   0.035 \\
  2437.0209 &  14.834 &  0.033 &   0.024 \\
  2437.0278 &  14.679 &  0.008 &   0.012 \\
  2437.0355 &  14.697 &  0.074 &   0.066 \\
  2633.3729 &  14.730 &  0.057 &  -0.088 \\
  2696.0850 &  14.778 &  0.100 &  -0.018 \\
  2696.1289 &  14.946 &  0.004 &  -0.006 \\
  2701.2598 &  14.527 &  0.013 &  -0.018 \\
  2701.2719 &  14.586 &  0.019 &   0.006 \\
  2701.3009 &  14.572 &  0.028 &  -0.052 \\
  2702.2897 &  14.803 &  0.017 &  -0.015 \\
  2708.2299 &  14.678 &  0.039 &  -0.029 \\
  2708.2428 &  14.420 &  0.034 &  -0.048 \\
  2708.2686 &  14.429 &  0.082 &   0.035 \\
  2708.2820 &  14.356 &  0.020 &  -0.029 \\
  2741.2177 &  14.580 &  0.078 &   0.019 \\
  2742.0505 &  14.651 &  0.048 &  -0.068 \\
  2742.0943 &  14.646 &  0.070 &  -0.099 \\
  2743.0436 &  14.595 &  0.030 &  -0.006 \\
  2743.0924 &  14.711 &  0.081 &  -0.019 \\
  2743.1037 &  14.457 &  0.094 &  -0.022 \\
  2746.0535 &  14.469 &  0.062 &   0.049 \\
  2746.0703 &  14.712 &  0.069 &  -0.092 \\
  2747.2671 &  14.472 &  0.046 &  -0.024 \\
  2752.2818 &  14.398 &  0.056 &  -0.010 \\
  2752.3235 &  14.309 &  0.055 &   0.066 \\
  2752.3354 &  14.496 &  0.074 &  -0.041 \\
  2757.2584 &  14.241 &  0.027 &  -0.038 \\
  2757.2833 &  14.270 &  0.006 &  -0.009 \\
  3017.2877 &  14.378 &  0.104 &   0.147 \\
  3017.3431 &  14.387 &  0.053 &   0.042 \\
  3018.3268 &  14.355 &  0.090 &  -0.036 \\
  3041.2464 &  14.469 &  0.024 &  -0.016 \\
  3041.2504 &  14.440 &  0.013 &   0.019 \\
  3091.1076 &  14.415 &  0.036 &   0.053 \\
  3091.1077 &  14.458 &  0.029 &   0.041 \\
  3091.1006 &  14.385 &  0.055 &   0.070 \\
  3091.1077 &  14.537 &  0.045 &   0.011 \\
  3093.2401 &  14.535 &  0.012 &   0.000 \\
  3094.1951 &  14.413 &  0.027 &   0.049 \\
  3094.1994 &  14.440 &  0.035 &   0.024 \\
  3100.1083 &  14.745 &  0.028 &  -0.054 \\
  3108.1097 &  14.459 &  0.098 &  -0.081 \\
  3108.1206 &  14.575 &  0.064 &  -0.013 \\
  3112.3084 &  14.416 &  0.065 &   0.004 \\
  3124.1235 &  14.515 &  0.056 &  -0.041 \\
  3124.1323 &  14.510 &  0.039 &  -0.042 \\
  3760.2646 &  14.261 &  0.013 &   0.025 \\
  3760.2955 &  14.570 &  0.042 &   0.060 \\
  3795.2234 &  13.896 &  0.047 &  -0.067 \\
  3795.2569 &  13.978 &  0.074 &  -0.104 \\
  3855.1142 &  14.204 &  0.008 &   0.012 \\
  3855.1228 &  14.499 &  0.068 &   0.097 \\
  3856.0373 &  14.568 &  0.063 &  -0.089 \\
  3856.0520 &  14.471 &  0.085 &  -0.121 \\
  3857.0662 &  14.425 &  0.000 &   0.000 \\
  3874.9808 &  14.190 &  0.039 &  -0.055 \\
  3886.0060 &  14.286 &  0.188 &  -0.267 \\
  3886.0105 &  14.106 &  0.016 &  -0.022 \\
  3886.0435 &  14.180 &  0.060 &  -0.085 \\
  3887.9957 &  14.200 &  0.066 &   0.094 \\
  3888.0086 &  14.187 &  0.127 &  -0.180 \\
  3888.0216 &  13.987 &  0.078 &   0.110 \\
  4231.7819 &  15.106 &  0.021 &   0.030 \\
  4231.8011 &  15.180 &  0.042 &  -0.059 \\
\enddata
\end{deluxetable}

\begin{table}
  \centering
  \begin{tabular}{llrc}\hline\hline
Period        & Amplitude     & Phase         & FAP \\
(d)           &               & ($\pi rad$)   &     \\
\hline
\multicolumn{4}{c}{V band data (1933--2004)} \\
$7022\pm1486 $  & $ 0.676\pm0.083$ &$ -0.81\pm0.12$  & $0.002$ \\
$3700\pm413  $  & $ 0.325\pm0.083$ &$ -1.00\pm0.26$  & $0.005$ \\
$1871\pm106  $  & $ 0.195\pm0.083$ &$  0.92\pm0.43$  & $0.006$ \\
$1297\pm51   $  & $ 0.218\pm0.083$ &$  0.40\pm0.38$  & $0.006$ \\
$847\pm22    $  & $ 0.321\pm0.083$ &$ -0.13\pm0.26$  & $0.005$ \\
$775\pm18    $  & $ 0.315\pm0.083$ &$  0.73\pm0.26$  & $0.005$ \\
$239\pm2     $  & $ 0.154\pm0.083$ &$ -0.99\pm0.54$  & $0.007$ \\
\hline
\multicolumn{4}{c}{Binned V band data (1933--2004)} \\
$13298\pm6298$  & $ 0.648\pm0.109$ &$ -0.79\pm0.17$  & $0.003$ \\
$8738\pm2720 $  & $ 0.598\pm0.109$ &$  0.41\pm0.18$  & $0.003$ \\
$3296\pm387  $  & $ 0.187\pm0.109$ &$ -0.69\pm0.58$  & $0.008$ \\
$1759\pm110  $  & $ 0.283\pm0.109$ &$ -0.70\pm0.38$  & $0.006$ \\
$947\pm32    $  & $ 0.240\pm0.109$ &$  0.86\pm0.45$  & $0.006$ \\
$775\pm21    $  & $ 0.313\pm0.109$ &$  0.93\pm0.40$  & $0.005$ \\
$233\pm2     $  & $ 0.149\pm0.109$ &$  0.21\pm0.58$  & $0.009$ \\
\hline
\multicolumn{4}{c}{V band data (1933--2004) with linear-trend removed} \\
$15268\pm10857$ & $ 0.687\pm0.109$ &$ -0.52\pm0.16$  & $0.002$ \\
$3026\pm427  $  & $ 0.218\pm0.109$ &$ -0.95\pm0.50$  & $0.007$ \\
$1961\pm179  $  & $ 0.224\pm0.109$ &$ -0.26\pm0.49$  & $0.007$ \\
$1282\pm77   $  & $ 0.276\pm0.109$ &$  0.11\pm0.40$  & $0.006$ \\
$1169\pm64   $  & $ 0.188\pm0.109$ &$  0.00\pm0.58$  & $0.007$ \\
$776\pm28    $  & $ 0.273\pm0.109$ &$ -0.11\pm0.40$  & $0.006$ \\
$642\pm19    $  & $ 0.207\pm0.109$ &$ -0.57\pm0.53$  & $0.007$ \\
\hline
\multicolumn{4}{c}{Post-1965 V band data} \\
$7203\pm1564 $  & $ 0.758\pm0.082$ &$ -0.19\pm0.11$  & $0.002$ \\
$3660\pm404  $  & $ 0.347\pm0.082$ &$ -0.59\pm0.24$  & $0.004$ \\
$1306\pm51   $  & $ 0.253\pm0.082$ &$  0.06\pm0.33$  & $0.005$ \\
$928\pm26    $  & $ 0.342\pm0.082$ &$ -0.91\pm0.24$  & $0.004$ \\
$847\pm22    $  & $ 0.332\pm0.082$ &$ -0.67\pm0.25$  & $0.004$ \\
$573\pm10    $  & $ 0.137\pm0.082$ &$ -0.11\pm0.61$  & $0.008$ \\
$238\pm2     $  & $ 0.127\pm0.082$ &$ -0.61\pm0.66$  & $0.008$ \\

\hline\hline
\end{tabular}
\caption{Periodic components given by the $CLEANest$ algorithm.}
\label{tab:tab01}
\end{table}

\clearpage

\begin{figure} \figurenum{1} \epsscale{0.5}
\plotone{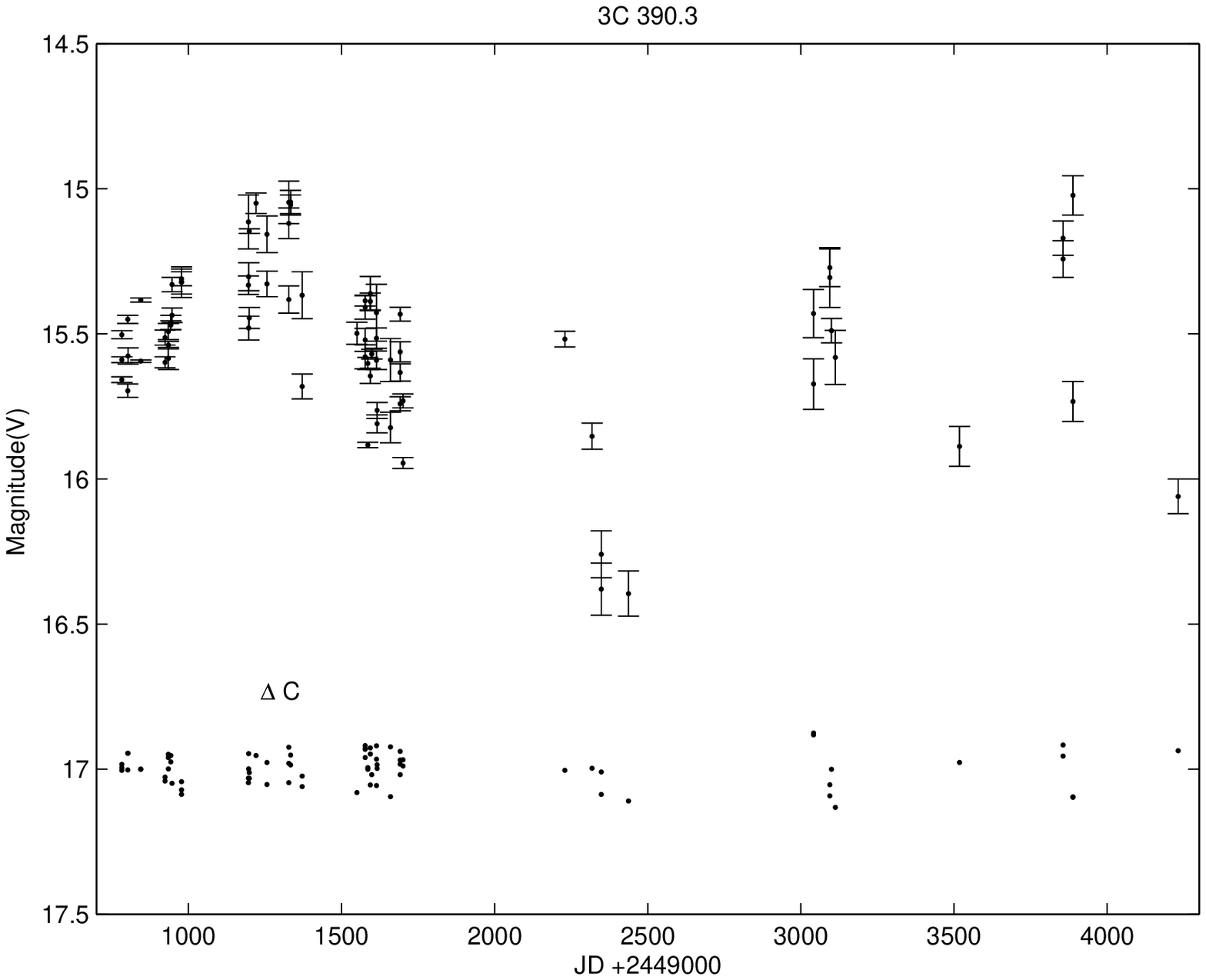}
 \caption{V-band light curve of 3C 390.3 (top) with the
 relative difference between comparison stars A and
B(bottom).} \label{fig1}\end{figure}

\clearpage

\begin{figure} \figurenum{2} \epsscale{0.5}
\plotone{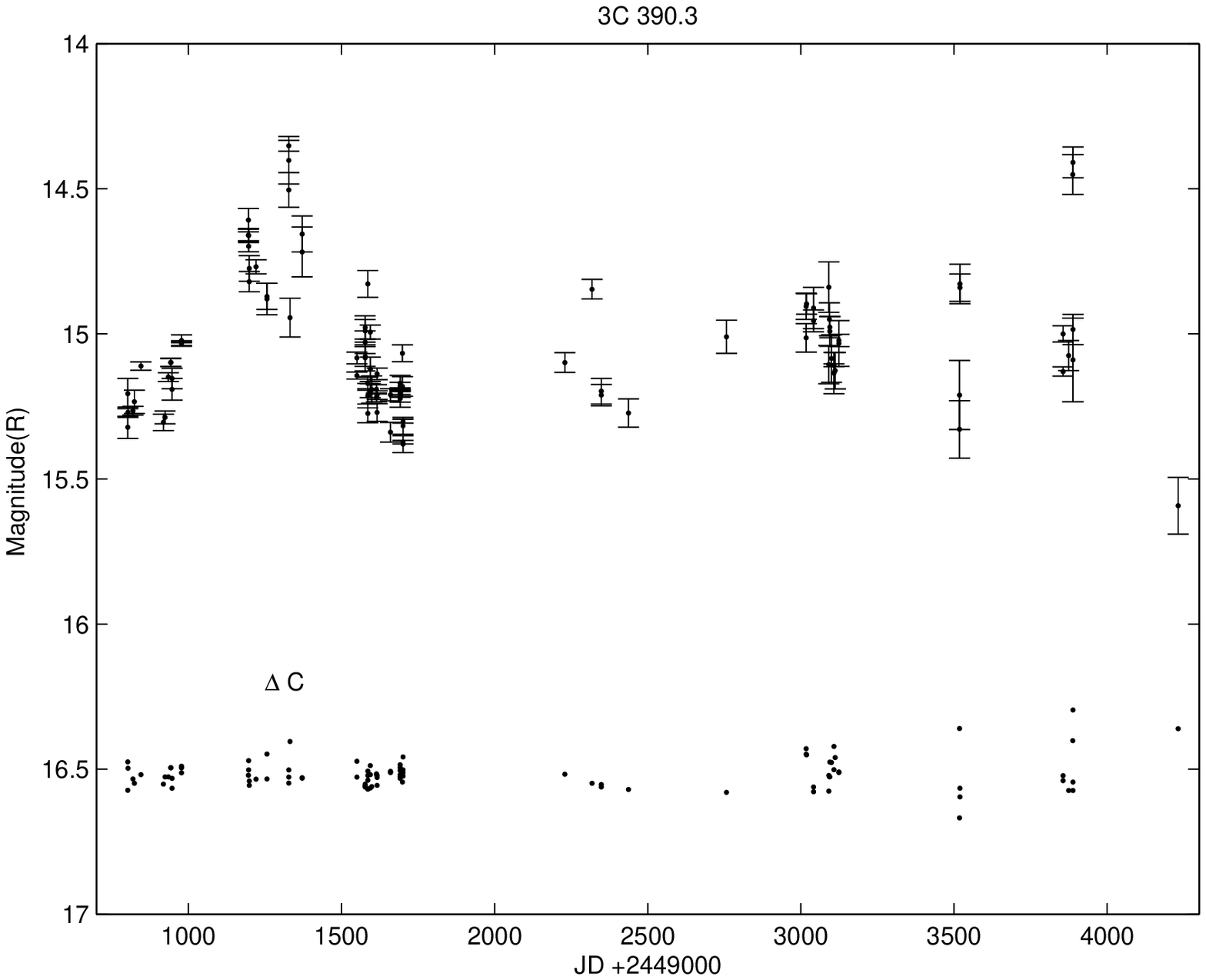} \caption{R-band Light curve of 3C 390.3 (top) with
the relative differences between comparison stars A and
B(bottom).}\label{fig2}\end{figure}

\clearpage

\begin{figure} \figurenum{3} \epsscale{0.5}
\plotone{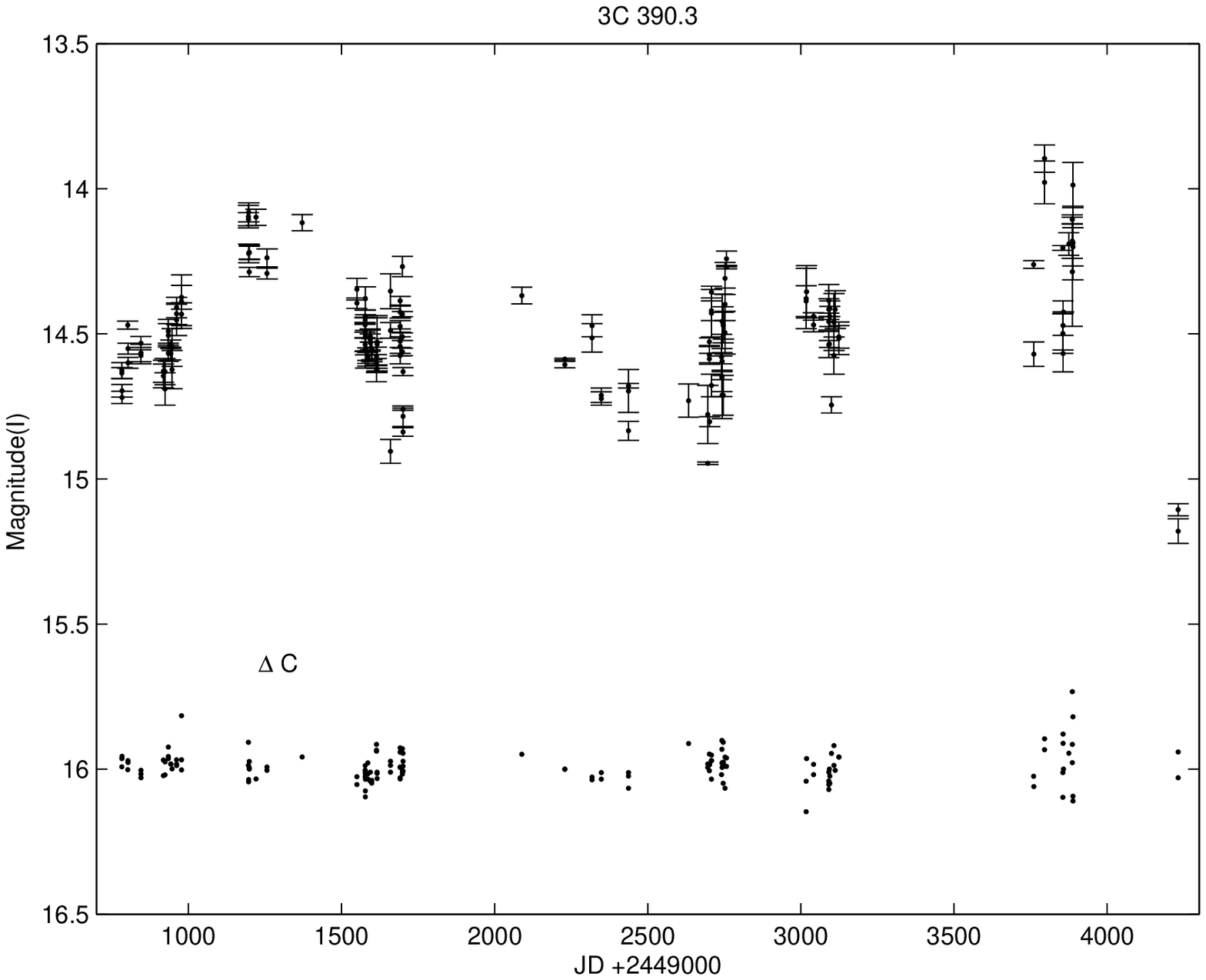} \caption{I-band light curve of 3C 390.3 (top) with
the relative differences between comparison stars A and
B(bottom).}\label{fig3}\end{figure}

\clearpage

\begin{figure}
\figurenum{4} \epsscale{0.5} \plotone{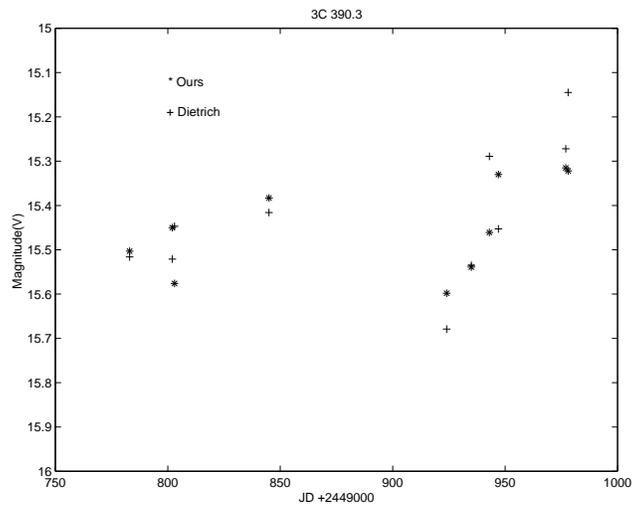} \caption{ Comparison
of the flux data given by Dietrich et al.(1998) with our V band
data. The Dietrich et al. fluxes have been converted to magnitudes
using the relationship $V=-2.5\times log(f)+17.961$.
}\label{fig4}\end{figure}

\clearpage

\begin{figure} \figurenum{5}
\epsscale{0.5} \plotone{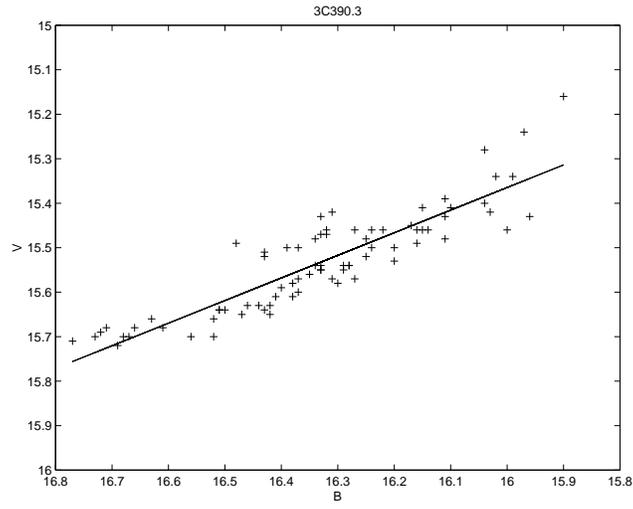} \caption{Comparison of B and V
magnitudes given by Shapovalova et al.
(2001).}\label{fig5}\end{figure}

\clearpage

\begin{figure} \figurenum{6}
\epsscale{0.5} \plotone{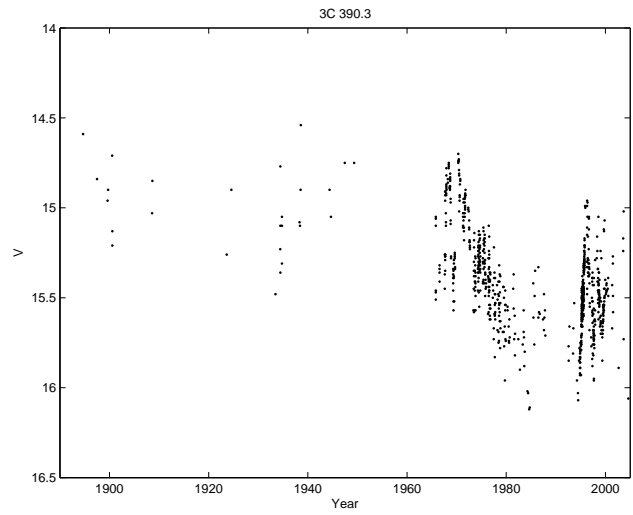} \caption{Reconstructed V-band
historical light curve of 3C 390.3.}\label{fig6}\end{figure}

\clearpage

\begin{figure}\figurenum{7}
  \begin{center}
    \resizebox{\hsize}{!}{
\includegraphics*{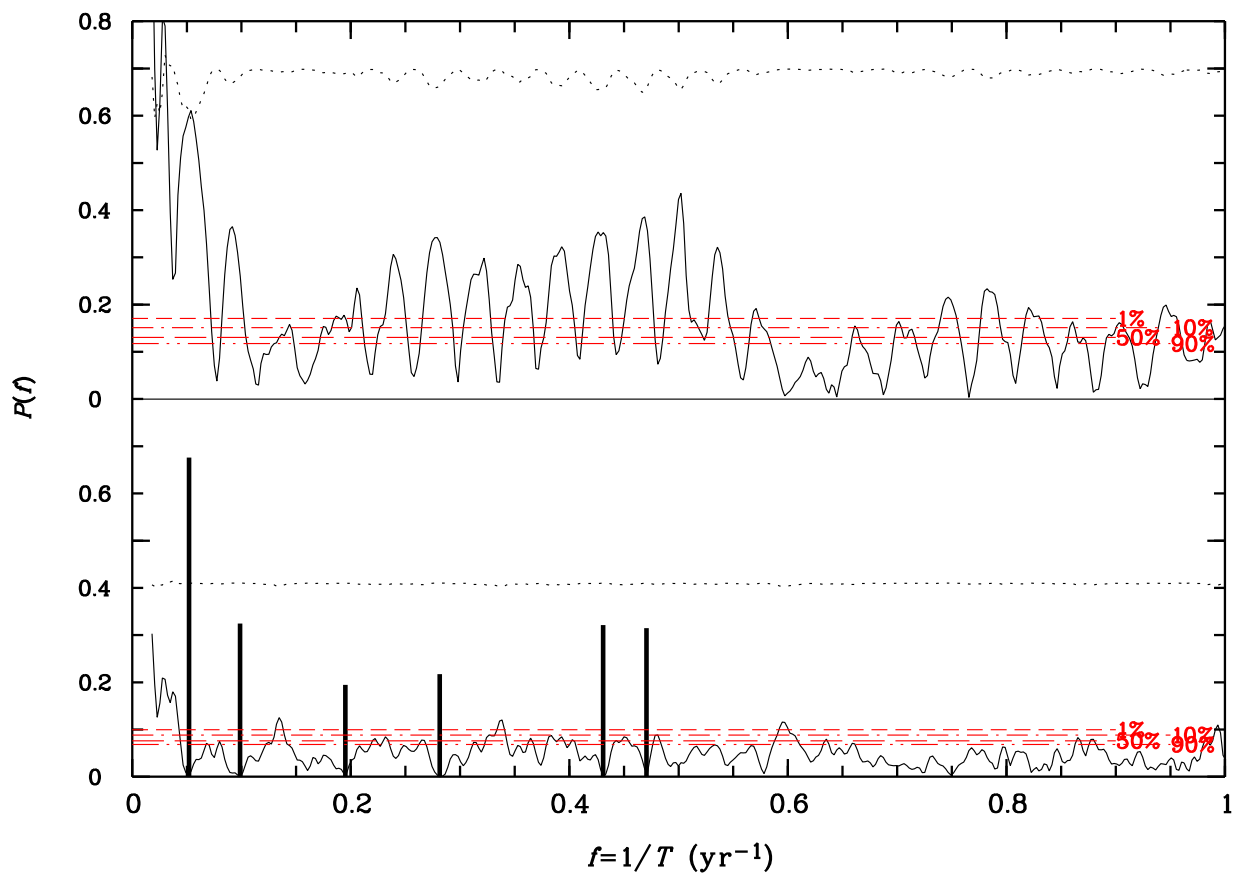}
    }
  \end{center}
  \caption{The Fourier power spectrum (top) and the $CLEANest$ spectrum (bottom) for
  the V-band light curve of 3C 390.2 from 1933-2004.  The dotted
  curves are the square deviations as a function of frequency. In the
  $CLEANest$ spectrum, six $CLEANest$ frequency components (thick vertical lines) and the
  residual spectrum are shown, one $CLEANest$ component with a frequency
  of 1.53 yr$^{-1}$ and an amplitude of 0.154 is off the figure.
  FAP significance levels of (from top to bottom) 0.01,0.10,0.50,0.90 are marked.}
  \label{fig7} \end{figure}

\clearpage

\begin{figure}[t]\figurenum{8}
  \begin{center}
    \resizebox{\hsize}{!}{
\includegraphics*{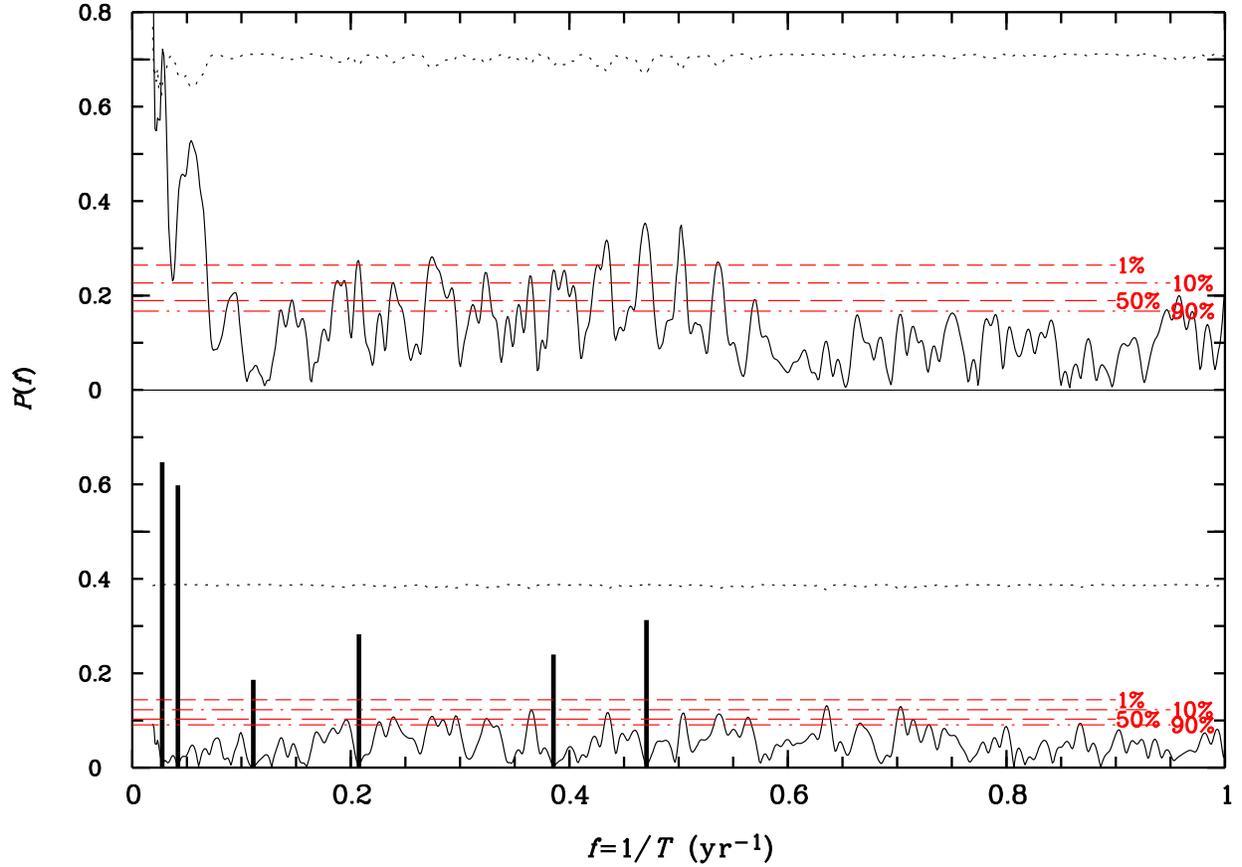}
}
  \end{center}
\caption{As for Fig. 7 but for the binned V-band light curve from
1933 to 2004. In the
  $CLEANest$ spectrum, six $CLEANest$ frequency components (thick vertical lines) and the
  residual spectrum are shown, one $CLEANest$ component with a frequency of 1.57
yr$^{-1}$, and an amplitude of 0.149 is off the figure.}
\label{h390f_dcdft} \label{8}
\end{figure}

\begin{figure}[t]\figurenum{9}
  \begin{center}
    \resizebox{\hsize}{!}{
    \includegraphics*{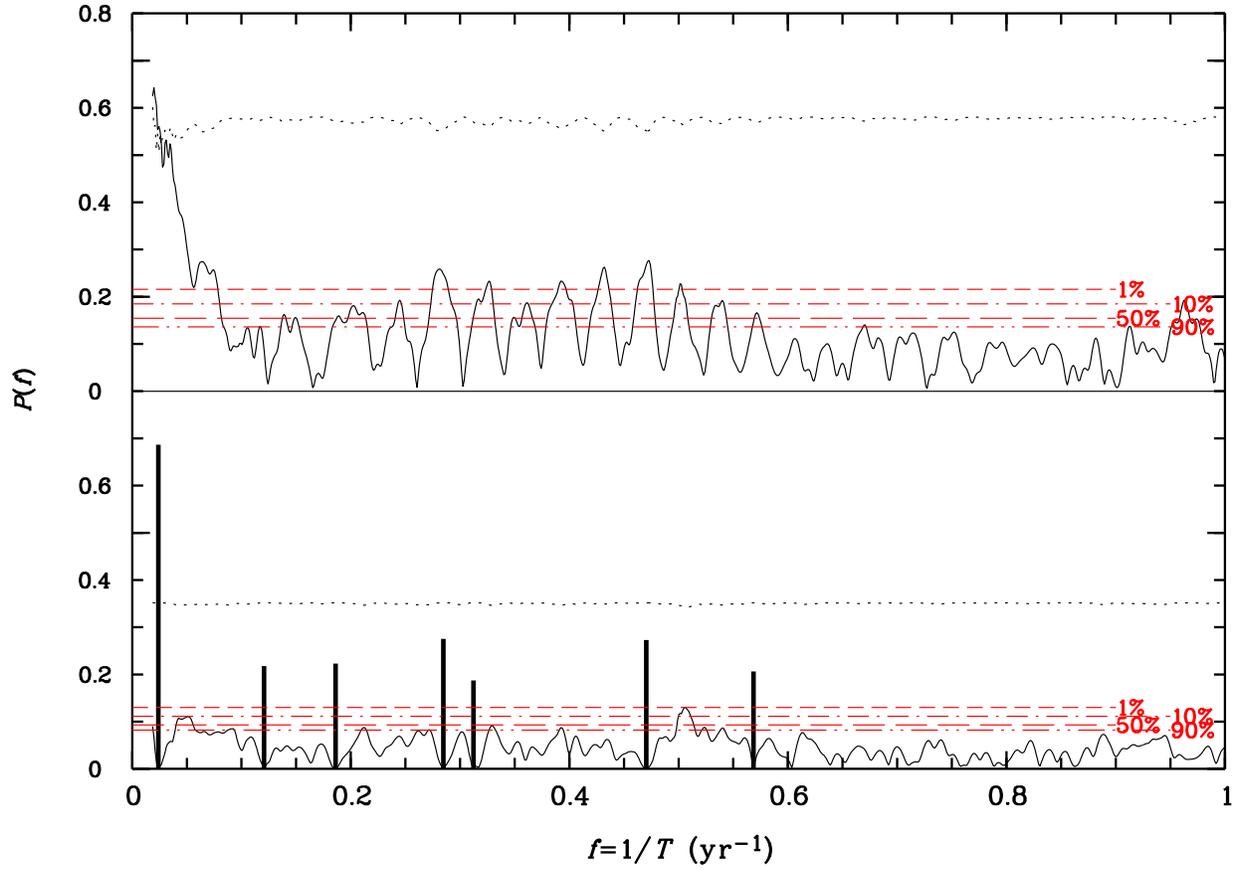}
    }
  \end{center}
\caption{As for Fig. 7 but with a linear-trend removed from the
light curve (see text). In the $CLEANest$ spectrum, seven $CLEANest$
frequency components are shown by thick vertical lines.}
\label{h390fm_dcdft} \label{9}
\end{figure}

\begin{figure}\figurenum{10}
  \begin{center}
    \resizebox{\hsize}{!}{
    \includegraphics*{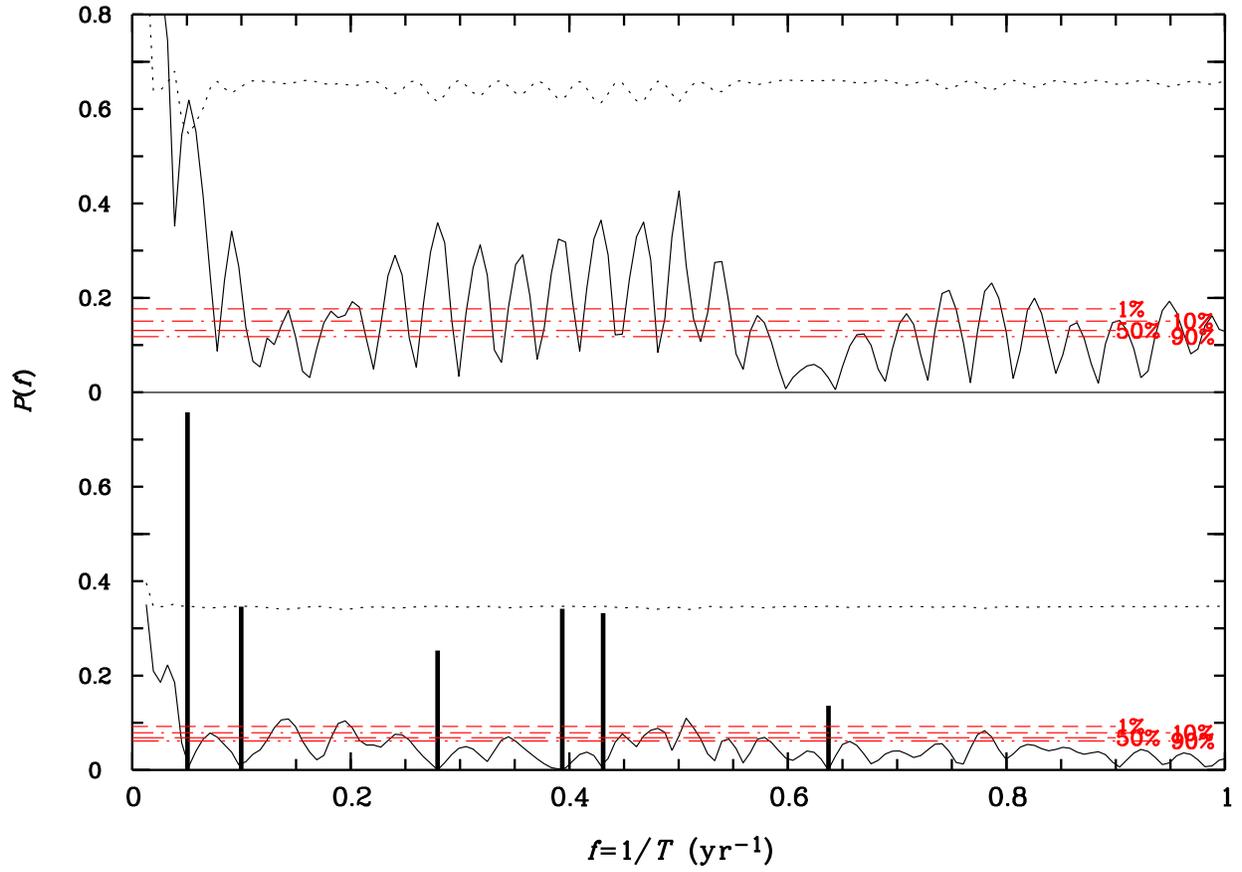}
    }
  \end{center}
\caption{As for Fig. 7 but only for the light curve from 1965 --
2004. One $CLEANest$ frequency components is beyond the range of the
figure.}
  \label{f390v_65_dcdft} \end{figure}

\clearpage

\begin{figure}[t]\figurenum{11}
  \begin{center}
    \resizebox{\hsize}{!}{
    \includegraphics*{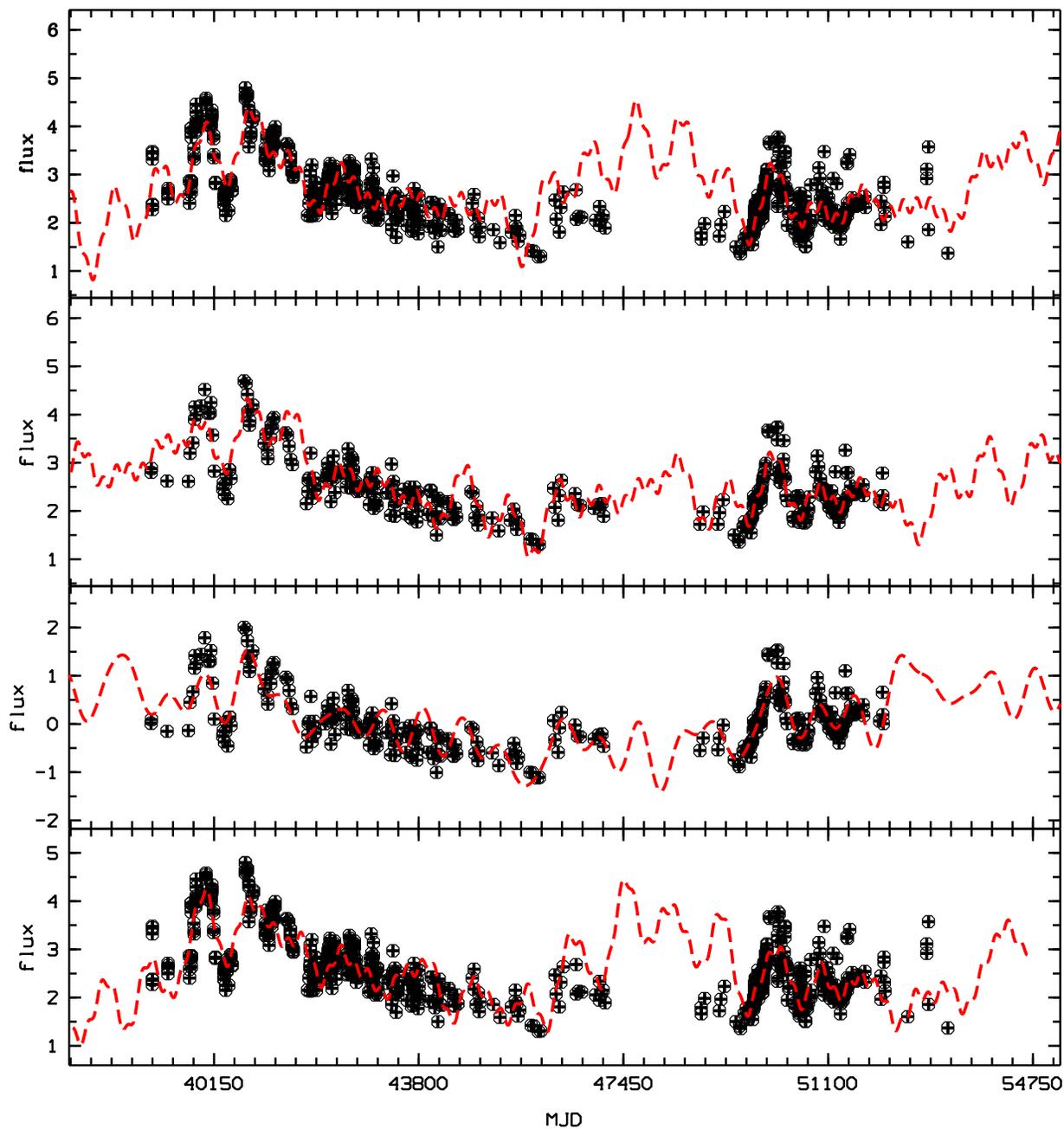}
    }
  \end{center}
\caption{Light curves with reconstructions (dashed curves) using
$CLEANest$ components. From top to bottom the panels show the
1933-2004 V band data, the binned data from the same period, the
data with a linear trend removed, and the post 1965 V band data. The
dashed curves are the theoretical results obtained by using
$CLEANest$ components listed in Table 4. For the top 3 panels, only
the post 1965 data are plotted for comparison.}
  \label{fig:fig01}
\label{11}\end{figure}

\end{document}